# Microclimatic variation in tropical canopies: A glimpse into the processes of community assembly in epiphytic bryophyte communities

Ting Shen[1,2,3] | Richard T. Corlett[3] | Flavien Collart[4] | Thibault Kasprzyk[2] | Xin-Lei Guo[5,6] | Jairo Patiño[7,8] | Yang Su[9] | Olivier J. Hardy[10] | Wen-Zhang Ma[11] | Jian Wang[12] | Yu-Mei Wei[13] | Lea Mouton[2] | Yuan Li[1,14] | Liang Song[1,15] | Alain Vanderpoorten[2]

[1]CAS Key Laboratory of Tropical Forest Ecology, Xishuangbanna Tropical Botanical Garden, Chinese Academy of Sciences, Menglun, Mengla, China; [2]Institute of Botany, University of Liège, Liège, Belgium; [3]Center for Integrative Conservation, Xishuangbanna Tropical Botanical Garden, Chinese Academy of Sciences, Menglun, Mengla, China; [4]Department of Ecology and Evolution (DEE), University of Lausanne, Lausanne, Switzerland; [5]Aba Academy of Ecological Protection and Development, Wenchuan, China; [6]State Key Laboratory of Environmental Geochemistry, Institute of Geochemistry, Chinese Academy of Sciences, Guiyang, China; [7]Island Ecology and Evolution Research Group, Instituto de Productos Naturales & Agrobiología (IPNA) – Consejo Superior de Investigaciones Científicas (CSIC), La Laguna, Spain; [8]Department of Botany, Ecology and Plant Physiology, University of La Laguna, La Laguna, Tenerife, Canary Islands, Spain; [9]UMR ECOSYS, INRAE AgroParisTech, Université Paris-Saclay, Thiverval-Grignon, France; [10]Evolutionary Biology and Ecology Unit CP 160/12, Faculté des Sciences, Université Libre de Bruxelles, Brussels, Belgium; [11]Key Laboratory for Plant Diversity and Biogeography of East Asia, Kunming Institute of Botany, Chinese Academy of Sciences, Kunming, China; [12]Bryology Laboratory, School of Life Science, East China Normal University, Shanghai, China; [13]Guangxi Key Laboratory of Plant Conservation and Restoration Ecology in Karst Terrain, Guangxi Institute of Botany, Guangxi Zhuang Autonomous Region and Chinese Academy of Sciences, Guilin, China; [14]School of Ecology and Environment, Hainan University, Haikou, Hainan, China and [15]Center of Plant Ecology, Core Botanical Gardens, Chinese Academy of Sciences, Menglun, Mengla, China

**Correspondence**
Ting Shen
Email: ting.shen@doct.uliege.be

Liang Song
Email: songliang@xtbg.ac.cn

**Funding information**
Candidates of the Young and Middle-Aged Academic Leaders of Yunnan Province, Grant/Award Number: 2019HB040; CAS 135 program, Grant/Award Number: 2017XTBG-F01 and 2017XTBG-F03; China Scholarship Council, Grant/Award Number: 201904910636; Fédération Wallonie-Bruxelles, Grant/Award Number: 1117545; Fonds De La Recherche Scientifique - FNRS, Grant/Award Number: 2.5020.11(Consortium des Equipements de Calcul Intensif); Ministerio de Ciencia e Innovación, Grant/Award Number: ASTERALIEN - PID2019-110538GA-I00 and RYC-2016-20506(Ramon y Cajal Program); National Natural Science Foundation of China, Grant/Award Number: 32171529; Natural Science Foundation of Yunnan Province, Grant/Award Number: 202101AT070059; Yunnan High Level Talents Special

## Abstract

1. Epiphytic communities offer an original framework to disentangle the contributions of environmental filters, biotic interactions and dispersal limitations to community structure at fine spatial scales. We determine here whether variations in light, microclimatic conditions and host tree size affect the variation in species composition and phylogenetic structure of epiphytic bryophyte communities, and hence, assess the contribution of environmental filtering, phylogenetic constraints and competition to community assembly.

2. A canopy crane giving access to 1.1 ha of tropical rainforest in Yunnan (China) was employed to record hourly light and microclimatic conditions from 54 dataloggers and epiphytic bryophyte communities from 408 plots. Generalized Dissimilarity Modelling was implemented to analyse the relationship between taxonomic and phylogenetic turnover among epiphytic communities, host-tree characteristics and microclimatic variation.

3. Within-tree vertical turnover of bryophyte communities was significantly about 30% higher than horizontal turnover among-trees. Thus, the sharp vertical variations in microclimatic conditions from tree base to canopy are more important than differences in age, reflecting the likelihood of colonization, area, and Support Plan, Grant/Award Number: YNWR-QNBJ-2020-066; Fundación BBVA, Grant/Award Number: INVASION-PR19_ECO_0046

**Handling Editor:** Glenn R Matlack

habitat conditions between young and old trees, in shaping the composition of epiphytic bryophyte communities.

4. Our models, to which microclimatic factors contributed most (83–98%), accounted for 33% and 18% of the variation in vertical turnover in mosses and liverworts, respectively. Phylogenetic turnover shifted from significantly negative or non-significant within communities to significantly positive among communities, and was slightly, but significantly, correlated with microclimatic variation. These patterns highlight the crucial role of microclimates in determining the composition and phylogenetic structure of epiphytic communities.

5. *Synthesis*. The mostly non-significant phylogenetic turnover observed within communities does not support the idea that competition plays an important role in epiphytic bryophytes. Instead, microclimatic variation is the main driver of community composition and phylogenetic structure, evidencing the role of phylogenetic niche conservatism in community assembly.

## 1 | INTRODUCTION

The relative influence of community assembly mechanisms varies depending on spatial and temporal scales (Kneitel & Chase, 2004; Kraft & Ackerly, 2014). In Grime's competitive, stress-tolerant, ruderal (CSR) theory of plant ecological strategies (Grime, 1977), community composition is controlled by selection for traits depending on levels of competition, stress and disturbance. Along a gradient of decreasing habitat filtering, community composition is expected to shift from a dominance of stress-tolerant species to competitive and ruderal species (Escobedo et al., 2021). At larger spatial scales, and hence, as variation in environmental conditions increases, community composition is conversely increasingly driven by environmental filtering (Powell et al., 2015). The contribution of environmental and biotic filters to community assembly is, however, often confounded (Cadotte & Tucker, 2017), especially at small spatial scales, at which both processes may occur (Xu et al., 2021).

In this context, phylogenetic turnover, which characterizes the phylogenetic structure of communities, offers an appealing framework to disentangle the processes involved in community assembly (Graham & Fine, 2008). While taxonomic turnover measures the extent to which some species are replaced by others along environmental gradients, phylogenetic turnover measures the extent to which species replacement is phylogenetically constrained, so that species within a community are more or less phylogenetically related to each other than expected by chance.

Positive phylogenetic turnover occurs when species in a community are more closely related to each other than species from different communities. A clumped phylogenetic distribution of taxa (phylogenetic clustering) indicates that habitat-use is a conserved trait within the pool of species in the community, and hence, evidences phylogenetic niche conservatism (Webb et al., 2002). The application of the phylogenetic niche conservatism hypothesis has substantial ecological and evolutionary implications because it makes it possible to determine whether niche preferences are evolutionarily labile or, to the reverse, are phylogenetically constrained, potentially hampering the chances of species to respond to climate change.

Negative phylogenetic turnover (phylogenetic overdispersion) occurs when species from the same community are more phylogenetically distant than species from different communities. While phylogenetic overdispersion points to non-random species assemblages, its interpretation has been controversial. In line with Darwin's competition-relatedness hypothesis, which posits that closely related species compete more strongly than distantly related ones (Cahill et al., 2008), phylogenetic overdispersion has primarily been interpreted in terms of competition among related species sharing limited resources within the same niche (Anacker & Strauss, 2014; Wiens & Graham, 2005). Phylogenetic overdispersion may, however, also result from niche convergence (Cavender-Bares et al., 2004) or facilitation (Valiente-Banuet & Verdu, 2007) among phylogenetically unrelated species.

Epiphytes appear as an interesting model to address the question of the factors shaping community structure at small spatial scales (Adams et al., 2017, 2019; Méndez-Castro et al., 2020). For epiphytes, host-trees typically function as habitat islands, exhibiting, like oceanic islands but at much smaller spatial scales and shorter time frames, sharp spatio-temporal variations in their abiotic environment (Adams et al., 2017; Hidasi-Neto et al., 2019; Itescu, 2019; Taylor & Burns, 2015).

The ecological conditions that prevail along a vertical gradient, from the base to the uppermost canopy, typically vary in terms of extrinsic (e.g. air humidity, light intensity, temperature) and intrinsic (physical properties of the substrate, such as bark texture and physico-chemistry, branch orientation and diameter) features (Cornelissen & ter Steege, 1989). In the outer canopy, the high light intensity and extremely low humidity, high wind exposure and daily variation in temperature and relative humidity, exert strong selection pressure for traits similar to those seen in desert habitats, such as leaf succulence, small stature, slow growth rate, water and nutrient storage capabilities and UV protection (Spicer & Woods, 2022). Progressing towards the tree base, physical stability of the support and relative humidity (RH) increases, while light, temperature and the daily variation in microclimatic conditions decrease, resulting in a more stable environment inhabited by species that are less tolerant of drought and high light intensity (Cornelissen & ter Steege, 1989; Freiberg, 1996; Watkins et al., 2007; Woods et al., 2015).

These conditions further vary along horizontal gradients not only due to differences among host-tree species in terms of branching architecture, bark texture and physico-chemistry (Hidasi-Neto et al., 2019) but also due to age differences among host trees. As the likelihood of colonization increases with time, old trees typically exhibit a higher epiphytic species richness than young ones (Taylor & Burns, 2015). Old trees also have a larger area for colonization and a higher diversity of micro-habitats than young ones (Paillet et al., 2019).

In this context, Grime's CSR theory of plant ecological strategies (Grime, 1977) allows us to make predictions on the importance of interactions among epiphytes depending on the ontogenetic stage of their host-tree and the habitat they occupy (Spicer & Woods, 2022). On a tree, competition is expected to increase from the canopy, characterized by large variations in light and microclimatic conditions, to tree base, with more buffered environmental variations. Competition is also expected to increase from young to old host-trees, as pioneer species progressively accumulate before entering competition with specialized competitors (Ellis & Ellis, 2013).

Despite these expectations, where and when competition and facilitation dominate, if at all, remains relatively unexplored in epiphyte ecology (Francisco et al., 2018; Spicer & Woods, 2022). Furthermore, while the vertical structures of epiphyte communities (Gehrig-Downie et al., 2013; Mota de Oliveira et al., 2009; Mota de Oliveira & ter Steege, 2015; Zotz, 2016) and, to a lesser extent, associated variations in microclimatic conditions (Murakami et al., 2022; Stuntz et al., 2002; Toivonen et al., 2017), have long been documented, no analysis has, to our knowledge, examined the relationship between microclimatic variation and species composition in a spatially explicit framework. In fact, although forest canopy science has been an active discipline since the 19th century, its progress has been slow, partly due to the limited accessibility of canopies (Nakamura et al., 2017) and the limited availability of fine-scale microclimatic data (De Frenne et al., 2021), a critical issue for canopy epiphytes (Murakami et al., 2022).

Bryophytes represent an important component of epiphytic floras, to which they contribute up to 75% of the biomass, and hence, play a key role in nutrient and water cycles (Gradstein et al., 2010). Bryophytes are poikilohydric and rely on rainfall or moisture in the atmosphere for water uptake. They are hence ideal models to investigate the impact of microclimatic variation on community composition, which strikingly varies from the base to the canopy (Mota de Oliveira & ter Steege, 2015; Sporn et al., 2010). Although mounting evidence points to the relevance of climatic niche conservatism for the assembly of bryophyte floras over large spatial and evolutionary time scales (Collart et al., 2021; Piatkowski & Shaw, 2019; Wilson & Coleman, 2022), whether shifts in community composition along vertical microclimatic gradients and along horizontal gradients in host-tree size are structured phylogenetically, that is, whether niche conservatism could operate at such micro-scales, remains to be tested.

Furthermore, it has been suggested that bryophytes may not compose communities similar to those of vascular plants, but instead, that the distributions of individual species would be driven by niche preferences and dispersal capacities, regardless of other species (Wilson et al., 1995). The unbounded relationship between epiphytic species richness and tree age has been interpreted in terms of the unrestricted increase in species richness in the absence of competition in unsaturated communities (Boudreault et al., 2000; Fritz, Brunet, & Caldiz, 2009). In line with this hypothesis, a significantly lower evenness, which could reflect weaker competition, was reported in bryophyte communities compared with those formed by vascular plants (Steel et al., 2004). Results from common garden experiments conversely revealed that competition is more important than temperature for the performance of bryophyte species (Greiser et al., 2021). The role of competition in bryophyte communities has, thus, long been questioned (Rydin, 2009). Wilson et al. (1995) concluded that there is community structure among bryophytes, in that species exclude each other to the same degree as higher plants do in their communities. They failed, however, to identify groups of species within a community that are mutually exclusive because of similarity in resource use, leading them to conclude that bryophyte species all form one guild.

Taking advantage of one of the world's 22 canopy cranes, the goal of the present study is to determine whether variations in light, microclimatic conditions and host tree size affect the variation in species composition and phylogenetic structure of epiphytic bryophyte communities, and hence, assess the contribution of environmental filtering, phylogenetic constraints and competition to community assembly. More precisely, we address the following questions: Is variation in species composition among epiphytic communities more important vertically, reflecting within-tree changes in microhabitat and microclimatic conditions, or horizontally, reflecting differences in age, and hence size and microhabitat diversity, among trees (Q1)? To what extent are these changes in community composition phylogenetically constrained (Q2)? Among communities, we test the hypothesis of an increasingly positive phylogenetic turnover along microclimatic gradients, pointing to phylogenetic niche conservatism (H1). Within communities, we test the hypotheses that species exhibit increasingly competitive interactions, and hence, increasing phylogenetic overdispersion, from the canopy to the base, and from young to old trees (H2).

## 2 | MATERIALS AND METHODS

### 2.1 | Study site and sampling design

This study took place in a 1.44 ha square plot in a lowland (643–700 m) seasonal rain forest (101°34′59.1″E, 21°37′2.6″N) in Mengla, one of the five subdistricts that together form Xishuangbanna National Natural Reserve (Yunnan, SW China). Mean monthly RH and mean monthly temperature recorded by dataloggers from 12 trees at 2 m during 2017–2019 were 95.3% and 20.8°C, respectively, with the coldest month in January (15.8°C) and the warmest month in June (25.2°C). This site was selected because it is equipped with an 81 m-high canopy crane (TCT7015-10E, Zoomlion Heavy Industry, Changsha, China) whose 60 m-long arm provides access to the canopy within a 1.1 ha circular area (Figure 1c).

Xishuangbanna National Natural Reserve covers an area of 242,510 ha that comprises the largest tropical forest area in China. The region experiences a typical monsoon climate with a 6-month dry season from November to April and a rainy season from May to October. In a 20-ha plot of tropical seasonal rainforest of Xishuangbanna, 468 tree species in 213 genera and 70 families were recorded. The tallest trees attain 70 m, and there are 4791.70 stems and a total basal area of 42.34 m$^2$ per hectare (Lan et al., 2012). Within the 1-ha plot investigated, the canopy layer (height > 30 m) is dominated by *Parashorea chinensis* (Dipterocarpaceae), which contributes 19.5% of the trees with a diameter at breast height (DBH) ≥ 5 cm and most of the tallest trees. It is accompanied by *Canarium album*, *Pometia tomentosa*, *Sloanea tomentosa* and *Semecarpus reticulata*. The sub-canopy layer (16–30 m) is dominated by *Ficus langkokensis*, *Litsea dilleniifolia*, *Barringtonia fusicarpa*, *Diospyros atrotricha* and *Pseuduvaria indochinensis*, and the understorey layer (6–16 m) by *Pittosporopsis kerrii*, *Baccaurea ramiflora*, *Diospyros xishuangbannaensis*, *Cleidion brevipetiolatum* and *Mitrephora maingayi*.

Epiphytic bryophytes were recorded only on the dominant host-tree species, *Parashorea chinensis*, to control for host specificity (González-Mancebo et al., 2003; Guan et al., 2017; Schmitt & Slack, 1990). *Parashorea chinensis* is an evergreen species, characterized by large buttresses, and hosts abundant epiphytic bryophytes (Shen et al., 2018). Although our analyses were restricted to the communities found on *P. chinensis*, these are representative of the entire epiphytic bryophyte community of the area. 102 epiphytic bryophyte species we found on 42 tree individuals, that is, slightly more than the 90 species reported from 69 individual trees belonging to 14 different tree species in the same plot in a previous investigation (Shen et al., 2018).

Trees with a DBH < 5 cm or covered by vines and lianas were discarded, resulting in a total of 42 with a DBH ranging from 5.4 to 135 cm. Each tree was divided into six height zones based on a slightly modified version of Johansson's (1974) zonation scheme (see e.g. Figure 7.11 in Zotz, 2016), which is not based on absolute height, but on tree architecture, as follows: tree base (zone 1), <2 m and corresponding to the buttresses; lower trunk (zone 2), between zone 1 and middle height of the trunk; upper trunk (zone 3), between the middle height of the trunk and the first ramifications of the canopy; inner, middle, and outer canopy (zones 4–6), corresponding to the lowest, middle and upper thirds of the canopy.

For each height zone, two plots were haphazardly located vertically. From the 504 initial plots, 96 had no bryophyte species, leading us to focus on 408 plots (see Shen (2021a) for individual plot coordinates) with at least one species, suitable for analyses of beta diversity (see below). Although orientation typically plays a limited role in explaining variation in epiphytic community composition in tropical cloud forests (Song et al., 2011), we controlled for this factor by sampling, for each plot, four sub-plots of 20 × 20 cm (as measured with a tape) on the trunk or branches (zones 1–4). These four sub-plots were organized in pairs, with the two plots of a pair being diametrically opposed and the pairs being perpendicular to each other. At zone 5, branches may be narrower than 20 cm, and we recorded epiphytes within a shape of 80 × 5 cm. At zone 6, we recorded an area of *c*. 400 cm$^2$ of twigs. This led to a total of 1632 sub-plots, 1156 of which had bryophytes.

Within each sub-plot, a complete species inventory was conducted. Representative specimens of each species were sampled in each sub-plot, resulting in 1156 collections that were subsequently analysed in the laboratory using relevant microscopic techniques and monographs (Shen et al., 2018). In some instances, the material available was too scanty to allow for an identification at the species level, and sometimes, even at the genus level. This was the case for six moss taxa, labelled as sp1–6, respectively (Table S4). Voucher specimens of each of the species included in the 1156 collections are kept at the Herbarium of the University of Liège (LG). The observations performed at the level of each sub-plot were then merged to produce presence-absence data for each of the moss and liverwort communities at the level of each plot (data available at https://doi.org/10.6084/m9.figshare.17057615.v8).

**FIGURE 1** Experimental design and 3D microclimatic modelling of temperature (T), relative humidity (RH), photosynthetically active radiation (PAR) and light intensity (L) in a 1.44 ha tropical canopy crane facility, Yunnan, SW China. (a) Vertical profile of day (orange line) and night (black line) monthly averages (and standard deviation, grey ribbon) of T, RH, PAR and L modelled at the level of Tree #1; (b) topographic map of the study area representing the position of the 42 sampled trees in a x-y space and the modelled horizontal variation in monthly average of day T, RH, PAR, and L at 2 m (n = 50) and at 50 m (n = 10) height, respectively; (c) experimental design. Circle diameters in (b) are proportional to tree DBH.

(a) Vertical microclimate #1

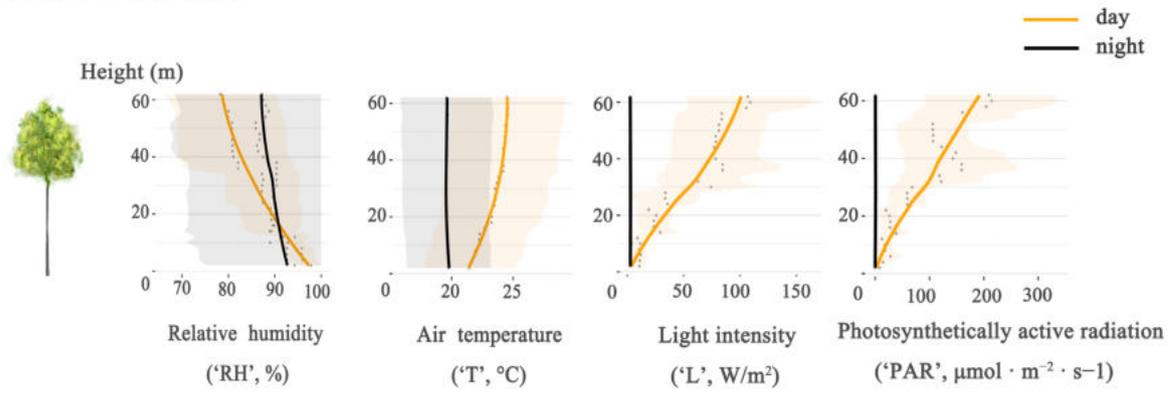

(b) Horizontal microclimate

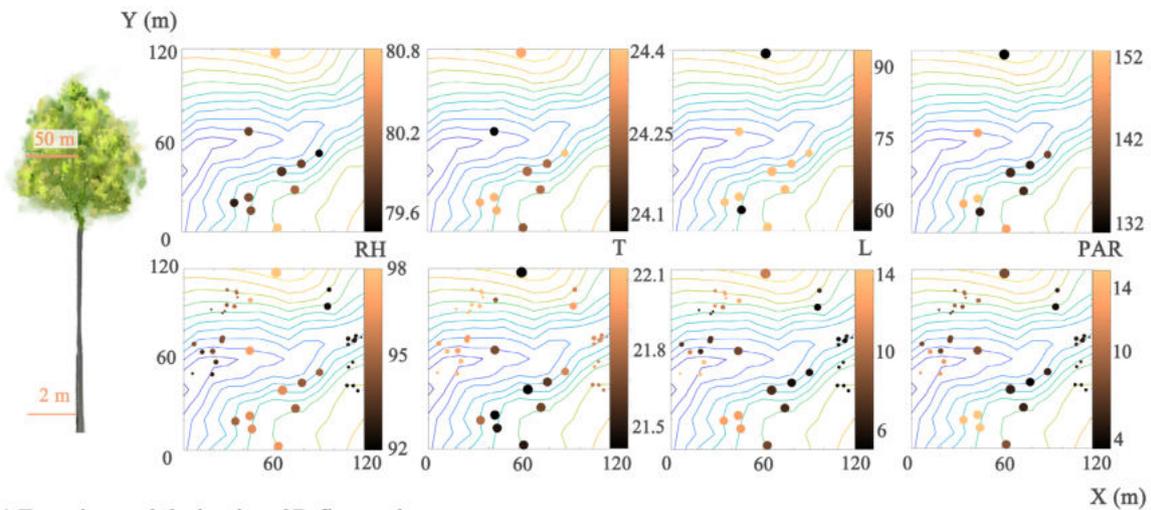

(c) Experimental design in a 3D fine scale

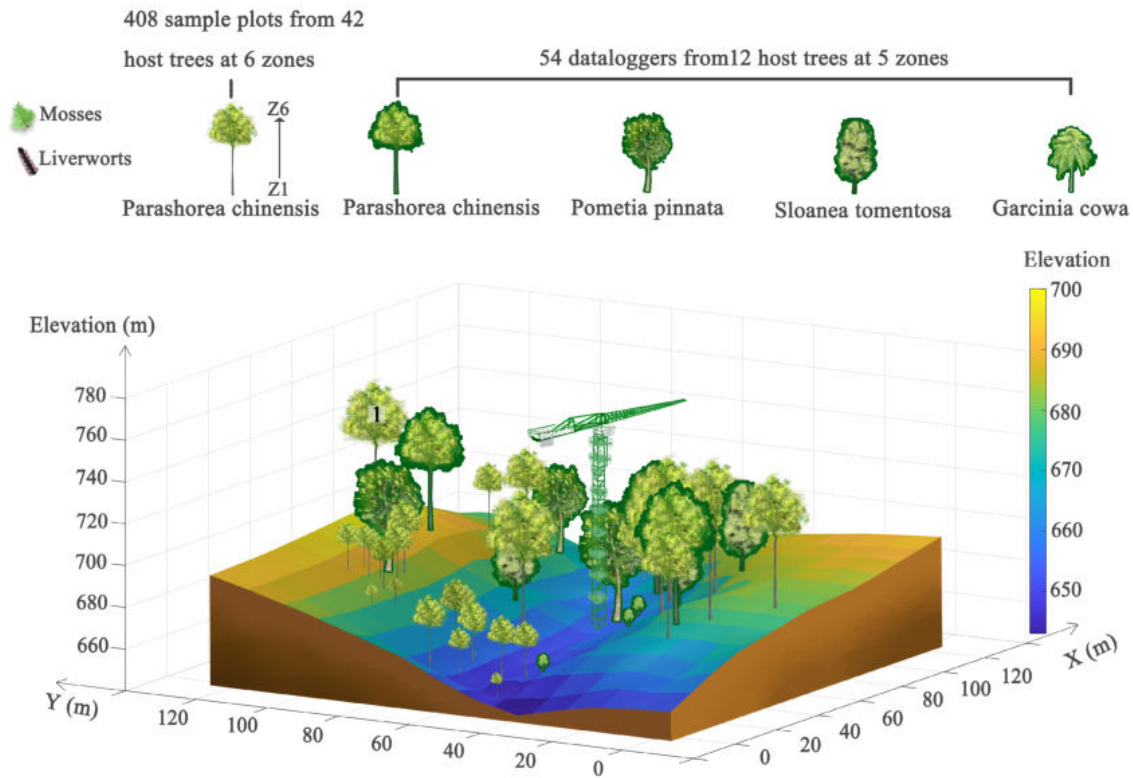

## 2.2 | Geographic and environmental variables

Nine ecological and geographic variables were recorded at each plot and were used to derive differences in ecological and geographic conditions among plots for subsequent analyses (Figure 2a).

The *X–Y* coordinates of each tree (Shen, 2021a) were used to compute the horizontal distance (hereafter, 'GeoDist', ranging between 0 and 114.82 m) among tree bases and the relative position between two trees in the *x–y* space (distance to a reference point, 'TreePos'). Tree height and plot height on the tree (Z coordinate) were measured with a tape from the hanging basket of the canopy crane. DBH of each tree was measured at 1.3 m above ground. The difference in DBH (hereafter, 'ΔDBH') was computed among all pairs of trees. We measured the elevation at 10 m intervals via the autopilot vehicle (LiAIR VUX-1350) equipped with VUX-1UAV Laser (RIEGL Laser Measurement Systems GmbH) and generated a 10 m resolution map with the measurements as pixel centroids using the RASTER package (Hijmans, 2021). The difference in elevation ('ΔElev') was then computed among all pairs of trees. To characterize the main habitat difference between the trunk and the canopy, a binary variable ('Treediv') was used to describe whether plots were both located on the trunk or in the canopy. In pairwise plot comparisons, two plots located on the trunk or the canopy had a 'Treediv' of 0, whereas pairs of plots including one plot from the trunk and the other plot from the canopy had a 'Treediv' of 1. Finally, we generated a binary variable indicating whether pairs of plots being compared are located on the same tree (0) or on different trees.

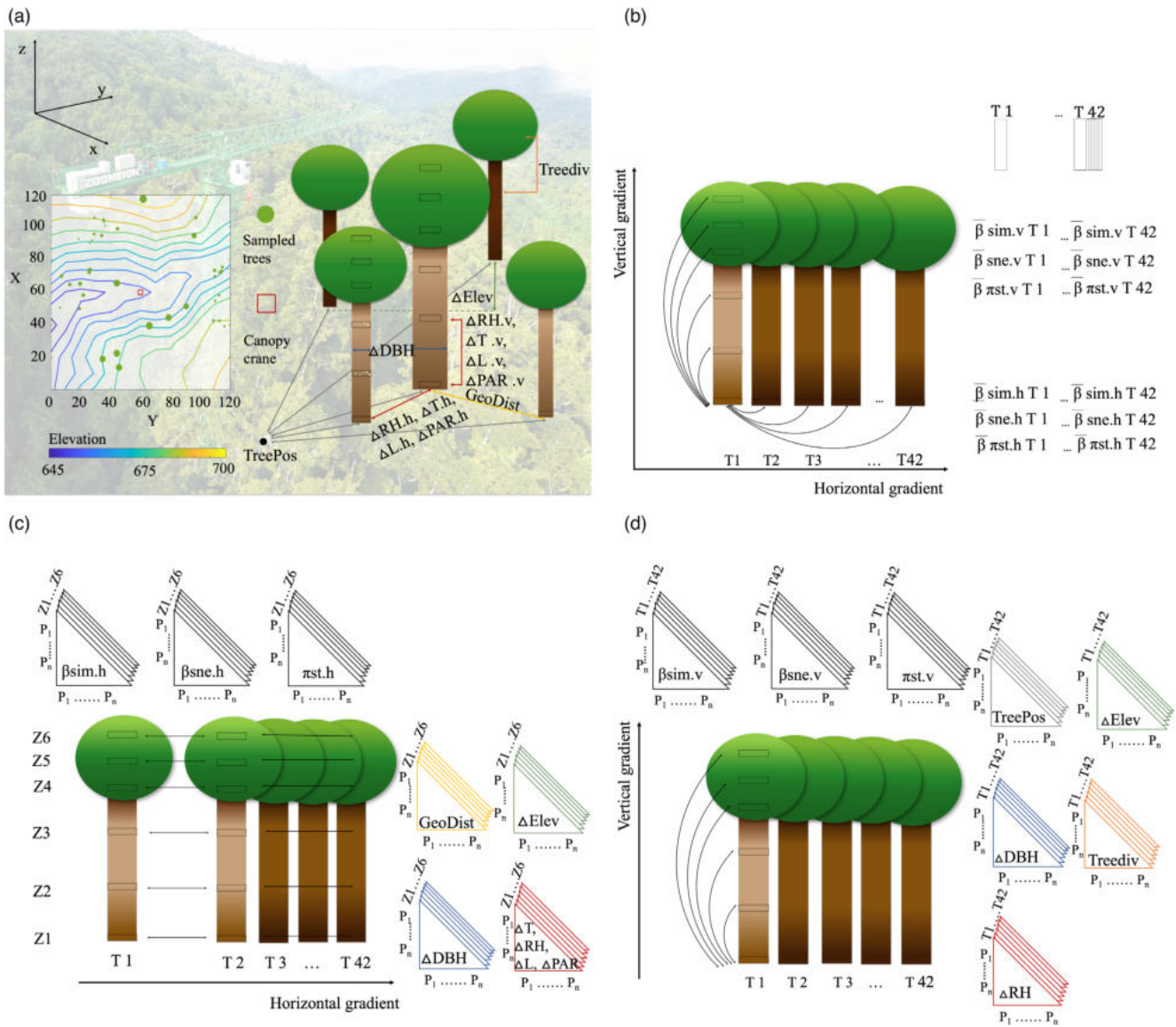

**FIGURE 2** Statistical design implemented for the analysis of the turnover (βsim) and nestedness (βsne) components of beta diversity and the phylogenetic turnover (πst) among epiphytic moss and liverwort communities. (a) Factors used in the analyses. Ecological and geographic distances among plots used as predictors include the horizontal (.h) and vertical (.v) differences in relative humidity (ΔRH), temperature (ΔT), light (ΔL), photosynthetically active radiation (ΔPAR), microtopography (relative difference in elevation among trees), ΔElev, derived from a topographic map of the area (insert), a variable accounting for within vs among trunk/canopy comparisons (Treediv), a variable reporting the relative position between two trees in the *x–y* space (distance to a reference point, TreePos), difference in DBH among trees (ΔDBH), the geographic distance among trees (GeoDist); (b) comparisons of the vertical (.v) and horizontal (.h) variation in βsim, βsne and πst; (c) horizontal variation in βsim, βsne and πst within each of the six height zones (Z1–Z6) as a function of differences in ΔRH, ΔT, ΔL, ΔPAR, ΔDBH and GeoDist; (d) vertical variation in βsim, βsne and πst among height zones within trees as a function of ΔRH, ΔDBH, ΔElev, TreePos and Treediv.

Air temperature ('T', °C), relative humidity ('RH', %), photosynthetically active radiation ('PAR', $\mu mol\,m^{-2}\,s^{-1}$) and light intensity ('L', $W\,m^{-2}$) were used to characterize light and microclimatic environmental conditions. T and RH were measured by HYS15 air temperature and relative Moisture Sensors, Unism, China. PAR and L were measured by LI-190R Quantum Sensor, LI-COR Biosciences (Figure S1). These variables were recorded by 54 dataloggers every hour during 30 months from July, 2017 to December, 2019, to document the spatio-temporal variation of microclimates and calibrate microclimatic models (see below). To cover the range of vertical and horizontal microclimatic variation within the 1.44 ha plot, these dataloggers were located at regular height intervals on 12 trees scattered across the study area (Table S1). Because of datalogger failures, for instance during storm events, data could not be collected by all the dataloggers over this entire time period. To avoid missing data, we therefore averaged the values recorded for the same hour and month across years (data available at https://doi.org/10.6084/m9.figshare.17057624).

## 2.3 | Spatial microclimate modelling

To predict the light and microclimatic conditions at each of the 408 plots from the data collected by the 54 dataloggers, we modelled hourly variation in T, L, RH and PAR in an X–Y–Z space (thus including tree height and elevation) using Random Forest (Liaw & Wiener, 2002) as implemented by the RANDOMFOREST package in R v4.0.4 (R Development Core Team, 2021). Random forest is an efficient technique to model complex interactions among predictor variables (Cutler et al., 2007) and non-linear responses (Arulmozhi et al., 2021), which has increasingly been used in climatic modelling (Arulmozhi et al., 2021; Ellis & Eaton, 2021; Su et al., 2021). 80% and 20% of the data were used to train and assess the models, respectively. The models were tuned by searching the best hyperparameter values after 10-fold cross-validation (see Figure S2 for a flow chart of the protocol used). Model predictions were used to compute the Euclidian distance (ΔT, ΔRH, ΔL, ΔPAR) of the hourly difference in predicted microclimatic conditions between each pair of plots.

## 2.4 | Taxonomic and phylogenetic beta diversity

Taxonomic beta diversity was partitioned into nestedness ($\beta_{sne}$) and turnover (here represented by Simpson's dissimilarity index, $\beta_{sim}$) with the BETAPART package (Baselga & Orme, 2012). Nestedness occurs when species found at the poorest plots represent a subset of the species pool found in the richest plots (Baselga, 2010), reflecting, for example, the accumulation of species on trees with time. Species turnover, in turn, reflects the shift in species composition that typically occurs along ecological gradients, and is expected here among communities from the base to the canopy.

Phylogenetic turnover was quantified through the $\pi_{st}$ statistics, which is a measure of the average phylogenetic distance among species within versus among plots (Hardy, 2008; Hardy et al., 2012). To determine whether there was a significant phylogenetic overdispersion ($\pi_{st} < 0$) or clustering ($\pi_{st} > 0$) of epiphytic communities (Q2), we computed an average $\pi_{st}$ from all pairwise comparisons of plots, both within and among height zones and DBH classes. We determined whether $\pi_{st}$ was significantly lower or higher than expected by chance by comparing the distribution of observed $\pi_{st}$ values with that obtained with 100 randomized phylogenies among the tips to build the distribution of the null hypothesis. For each of the 100 randomly resolved phylogenetic trees, we re-computed the pairwise $\pi_{st}$ values among plots, which served to generate the distribution of 100 average $\pi_{st}$ among plots that would be expected if phylogenetic relationships among species were random. An observed average $\pi_{st}$ was significantly lower or higher than expected by chance if it was lower or higher than 95% of the values obtained after phylogeny permutations.

Phylogenetic distances among species pairs were computed from the moss and liverwort chronograms produced by Laenen et al. (2014). These chronograms resulted from large-scale analyses using genera as sampling units and including a single species per genus. The liverwort phylogeny was derived from the analysis of eight genes from all genomic compartments and includes 303 genera, representing 84% of the total extant generic diversity. The moss phylogeny was based on the analysis of one nuclear, one mitochondrial and one chloroplast gene and includes genera representing 64% of the total extant generic diversity of mosses. Phylogenetic trees were pruned to only keep the tips corresponding to observed species to generate suitable distributions of the null hypothesis (Hardy & Senterre, 2007). Twelve genera, which were not sampled in the phylogenies, were assigned to their closest genus based on phylogenetic evidence (Table S2). Since the phylogenies included a single species per genus, all congeneric species included in the present dataset were grafted onto the genus-level phylogeny, ensuring that phylogenetic relationships and branch lengths within genera were random and that the ages of genus crown nodes ranged between time present and the age of their stem node. In total, 100 trees with randomly resolved relationships among congeneric species were generated and separately analysed to take phylogenetic uncertainty into account. Taxa which could not be identified at the genus level were omitted from the analysis.

## 2.5 | Statistical analyses

Comparing vertical and horizontal patterns in $\beta_{sim}$, $\beta_{sne}$ and $\pi_{st}$ (Q1) involves the inclusion of the same plot multiple times, violating the assumption that the observations are independent from each other. We therefore computed, for each tree, the average βsim, βsne and πst among plots on the same tree, avoiding comparisons among plots located within the same height zone. This generated a distribution of 42 average vertical βsim, βsne and πst (Figure 2b). We then computed, again for each tree, the average βsim, βsne and πst among plots between the focal tree and all other trees, making sure to restrict the comparisons among plots from the same height zone and to trees that belong to the same DBH class to avoid non-homologous comparisons (e.g. plots from the canopy of a 9 m and 70 m tree). The categories considered were small trees (DBH of 5.4–19.6 cm, $n = 16$), medium trees (DBH of 20.3–39.6 cm, $n = 15$), and large trees (DBH of 66.9–135 cm, $n = 11$). This generated 42 average horizontal βsim, βsne and πst distributions (Figure 2b). The vertical and horizontal distributions of average βsim, βsne and πst significantly departed from normality (Kolmogorov–Smirnov test, $p < 0.001$) and homoscedasticity (Bartlett's test, $p < 0.001$) for both mosses and liverworts. We therefore applied a paired Wilcoxon rank test to test the hypothesis that, on average, vertical βsim, βsne and πst are larger than horizontal βsim, βsne and πst values.

To disentangle the contribution of the factors affecting βsim and βsne along horizontal and vertical gradients, we implemented Generalized Dissimilarity Modelling (GDM; Ferrier et al., 2007). Because the GDM program needs values in the biological dissimilarity matrix ranging between 0 and 1, πst values were rescaled accordingly in these analyses. For horizontal gradients, we generated six matrices (Figure 2c), each of which encompassed all pairwise comparisons among plots located within the same height zone within and among trees. Predictors included GeoDist, ΔDBH, ΔElev, ΔT, ΔRH, ΔL and ΔPAR among each pair of plots. For the vertical patterns, we focused on pairs of plots located on the same tree and generated a matrix including all pairwise plot comparisons within the 42 trees (Figure 2d). Predictors included ΔT, ΔRH, ΔL, ΔPAR and Treediv. To inform the model of the structure in the data, wherein only within-tree comparisons were allowed, we added the variable TreePos.

To circumvent collinearity among predictors, we computed, for each analysis, the correlation between environmental predictors as well as the variation inflation factor (VIF). If any of the predictors exhibited a VIF >5, the predictor with the highest VIF was removed. The VIF of the remaining variables was re-computed, and so on until all predictors had a VIF <5 (Akinwande et al., 2015). We then performed variable significance testing with 50 permutations per step until only significant ($p < 0.05$) variables remained in the model. We finally estimated the contribution of each variable to the model using the gdm.varImp function (Fitzpatrick et al., 2021).

To determine how πst varies along environmental gradients, we performed analyses at the level of average πst within and among communities and pairwise πst among plots. We visualized the variation of average πst per DBH class and height zone as a function of an ordinal ecological distance, computed as the number of height zone difference between communities. The significance and strength of this relationship was assessed with a Mantel test (VEGAN package, Oksanen et al., 2020). We then performed a second series of analyses using pairwise plot comparisons using the GDM framework described above.

## 3 | RESULTS

### 3.1 | Microclimatic modelling

Microclimatic conditions exhibited substantial vertical variations (Figure 1a). Between 2 and 62 m above ground, day (8 am–7 pm) RH ranged between 53.6% and 99.9% (monthly average 79.7–93.5%), day temperature between 12.0 and 31.7°C (monthly average 17.8–27.8°C), light intensity between 2.3 W m$^{-2}$ and 208.0 W m$^{-2}$ (monthly average 27.8–51.0 W m$^{-2}$), and PAR between 0.0 μmol m$^{-2}$ s$^{-1}$ and 407.6 μmol m$^{-2}$ s$^{-1}$ (monthly average 41.4–94.0 μmol m$^{-2}$ s$^{-1}$). Average variations ± standard deviation (SD) in the day between 2 m and 50 m were of 2.4±1.6°C for temperature, −16.7±10.2% for RH, 85.0±47.3 W m$^{-2}$ for light intensity, and 128.0±77.1 μmol m$^{-2}$ s$^{-1}$ for PAR. Horizontal variation was more subtle (Figure 1b), with average maximum variations (differences between maximum and minimum) in the day at 2 m and at 50 m reaching, respectively, 0.9±0.5°C and 0.5±0.3°C for temperature, 4.4±1.9% at and 3.3±2.1% for RH, 13.1±13.5 W m$^{-2}$ and 24.2±18.0 W m$^{-2}$ for light, 12.7±12.4 μmol m$^{-2}$ s$^{-1}$ and 48.9±40.4 μmol m$^{-2}$ s$^{-1}$ for PAR. This variation was captured by Random Forest models, with $R^2$ ranging from 0.96 for PAR to 0.99 for temperature (Figure S3; Table S3).

### 3.2 | Species richness and composition

Totals of 50 moss and 52 liverwort species were recorded (Table S4). The base (zone 1) was dominated by *Circulifolium microdendron*, *Caduciella mariei* and *Claopodium prionophyllum*. Along the trunk (zones 2–3), *Plagiochila parviramifera* and *Plagiochila fordiana* prevailed on small trees, and *Frullania monocera*, *Mastigolejeunea repleta* and *Caduciella mariei* on large trees. The most representative species were *Erythrodontium julaceum* and *Groutiella tomentosa* in the inner canopy of large trees, *Lejeunea flava*, *Cheilolejeunea eximia* and *Groutiella tomentosa* in the inner canopy of medium and large trees, and *Frullania ericoides*, *Acrolejeunea recurvata* and *Sematophyllum subhumile* in the outer canopy of large trees. The most frequent epiphylls were *Caudalejeunea reniloba*, *Cololejeunea planissima* and *Leptolejeunea subacuta*.

In mosses, species richness decreased from the base, with an average±(SD) of 4.5±1.5 species per plot (14.5±3.3 species per DBH class) to 2.6±1.0 species per plot (4.5±2.5 species per DBH class) in the outer canopy of large trees. In liverworts in contrast, species richness increased from the base (small trees), with 1.4±0.5 species per plot (7.6±2.7 species per DBH class), to the outer canopy, with 2.9±1.3 species per plot (12.3±6.9 species per DBH class) on large trees (Figure 3).

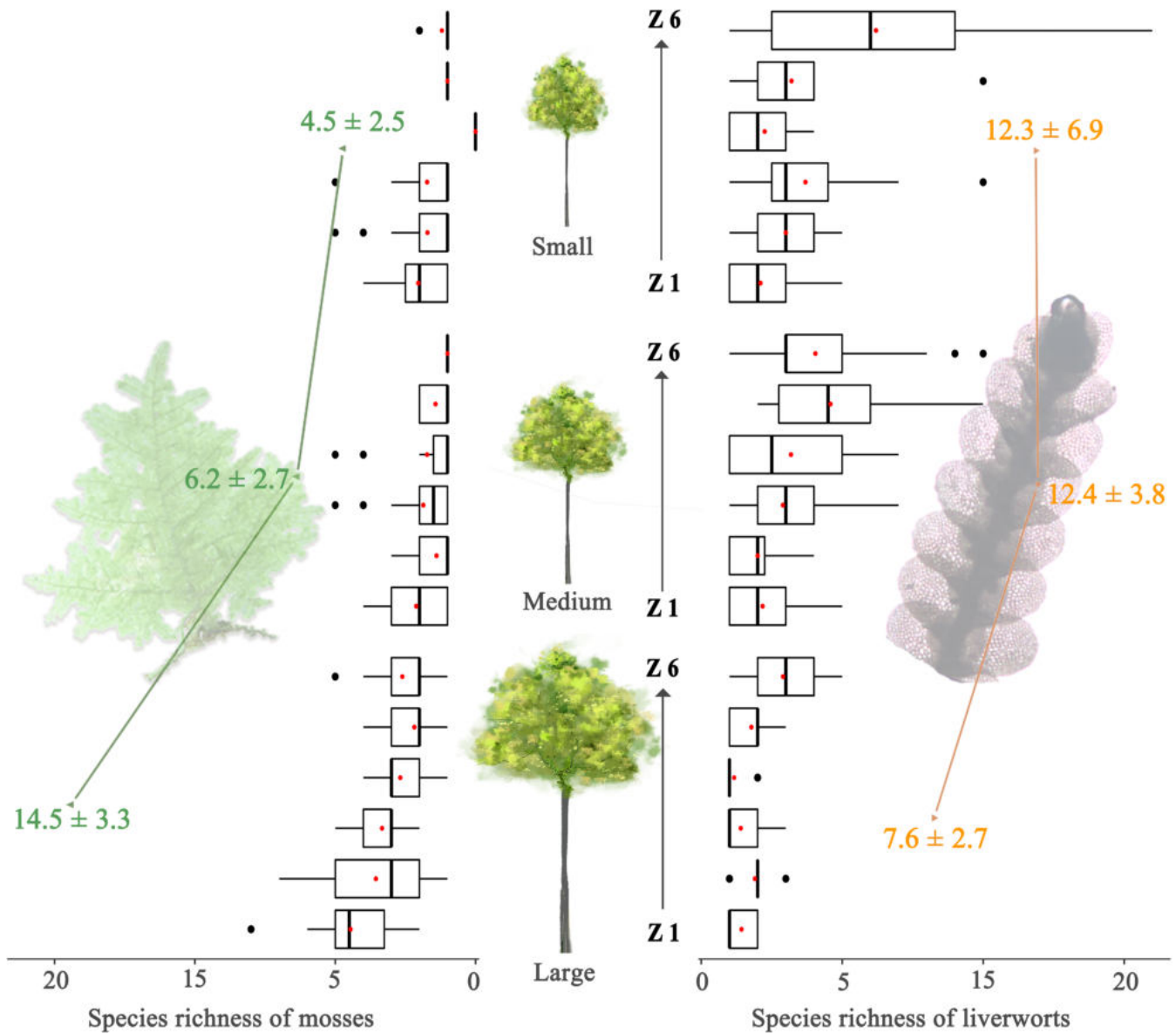

**FIGURE 3** Vertical variation in species richness of epiphytic mosses (left) and liverworts (right) in a 1.44 ha tropical canopy crane facility, Yunnan, SW China. The box-plots [showing the first and third quartiles (upper and lower bounds), second quartile (center), average (red dots), 1.5* interquartile range (whiskers) and minima–maxima beyond the whiskers] represent species richness per height zone on small (DBH of 5.4–19.6 cm, $n = 16$), medium (DBH of 20.3–39.6 cm, $n = 15$) and large (DBH of 66.9–135.0 cm, $n = 11$) *Parashorea chinensis* individuals. The line represents the average (mean ± SD) of moss (green triangles) and liverwort (orange triangles) species richness per DBH class.

## 3.3 | Taxonomic beta diversity

Turnover contributed about three- to fivefold more to taxonomic beta diversity than nestedness (Figure 4). Vertical turnover of moss and liverwort communities was significantly higher than horizontal turnover (average ± SD of vertical $\beta sim = 0.59 \pm 0.25$ and $0.66 \pm 0.24$, average horizontal $\beta sim = 0.42 \pm 0.20$ and $0.50 \pm 0.16$ in mosses and liverworts, respectively, $p < 0.001$ for the differences between vertical and horizontal $\beta sim$ in both mosses and liverworts). Nestedness exhibited the reverse pattern (average horizontal $\beta sne = 0.15 \pm 0.07$ and $0.14 \pm 0.06$, average vertical $\beta sne = 0.11 \pm 0.09$ and $0.14 \pm 0.06$ in mosses and liverworts, respectively, $p < 0.001$ for the differences between vertical and horizontal $\beta sne$ in both mosses and liverworts).

In GDM analyses focusing on the horizontal variation in taxonomic turnover within the same height zone, which accounted, on average, for $19.3 \pm 21.8\%$ (mosses) and $11.1 \pm 2.6\%$ (liverworts) of the explained deviance across height zones, the difference in DBH among trees was the best predictor, with a relative contribution ranging between 64.7% and 99.8% across height zones (Table S5). For horizontal nestedness, no model was significant except for liverworts in height zone 2 (Table S5).

In analyses focusing on the vertical variation in beta diversity (Figure S4), models contributed to 33.3% and 17.8% of the total deviance of species turnover in mosses and liverworts, respectively. Difference in RH among plots was the most important variable in the model, contributing to 98.8% and 83.0% of the deviance in species turnover of mosses and liverworts, respectively, while Treediv contributed to less than 1% and 15%. No model was significant for the vertical variation in nestedness.

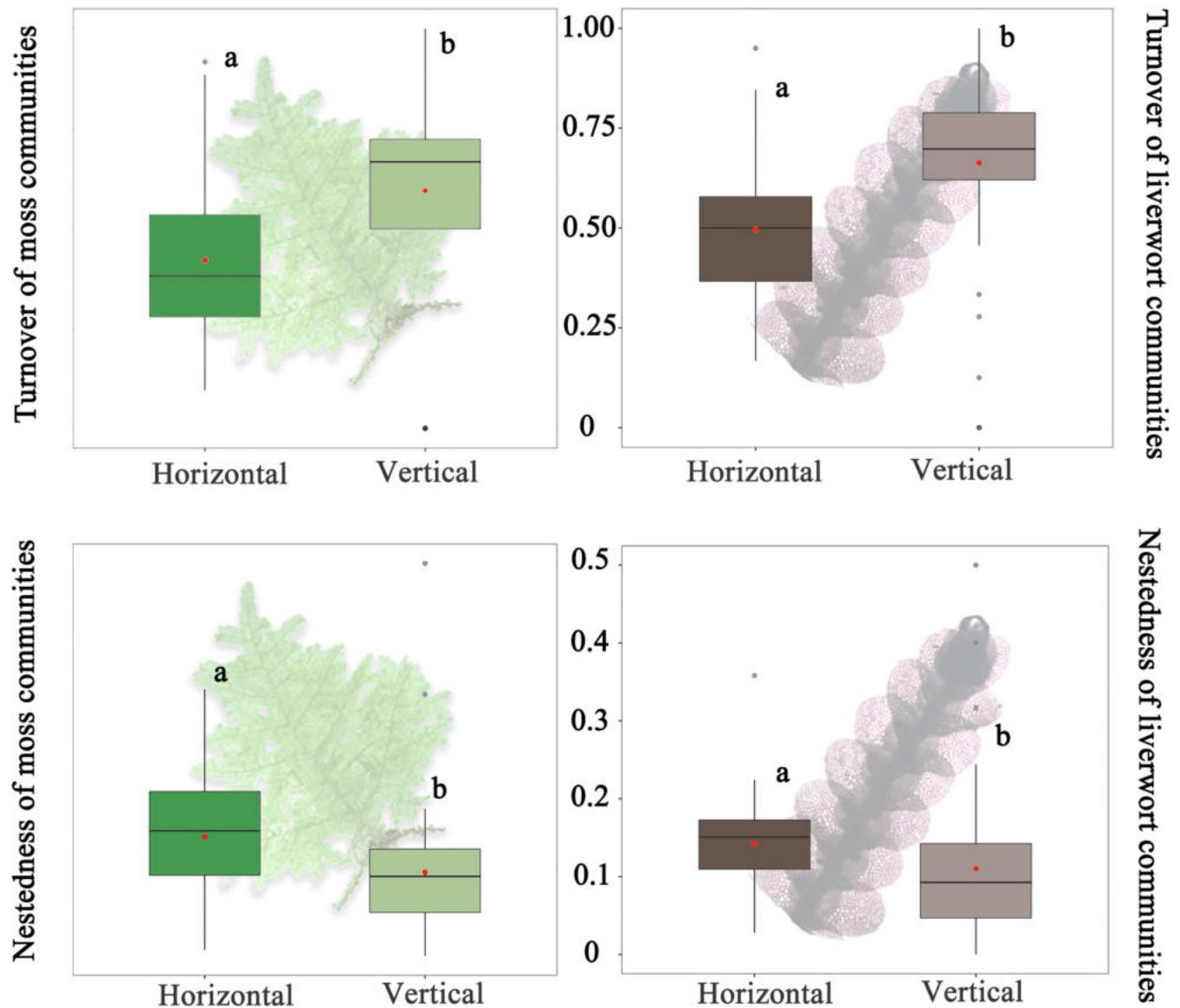

**FIGURE 4** Vertical and horizontal patterns of turnover and nestedness in epiphytic mosses and liverworts on *Parashorea chinensis* in a 1.44 ha tropical canopy crane facility, Yunnan, SW China. Box-plots [showing the first and third quartiles (upper and lower bounds), second quartile (center), average (red dots), 1.5* interquartile range (whiskers) and minima-maxima beyond the whiskers] represent the vertical turnover, nestedness for pairs of plots on the same tree and horizontal turnover, nestedness within the same height zone and among trees belonging to the same class of diameter at breast height (DBH; small trees, DBH of 5.4–19.6 cm, medium trees, DBH of 20.3–39.6 cm, and large trees, DBH of 66.9–135 cm) of moss and liverwort epiphytic communities. Letters above each box-plot indicate which comparisons significantly differ.

## 3.4 | Phylogenetic turnover

Average πst per height zone and DBH class significantly increased ($r = 0.25$, $p < 0.001$ for liverworts and $r = 0.09$, $p < 0.001$ in mosses) and shifted from mostly non-significant or rarely significantly negative (at the base of small trees in both mosses and liverworts and in the outer canopy of large trees in mosses, Table S6) to consistently significantly positive along a gradient of height zone differences (Figure 5).

In pairwise plot comparisons, horizontal phylogenetic turnover could not or could marginally be predicted from horizontal variation in microclimatic conditions and differences in DBH among trees (Table S5). For vertical phylogenetic turnover (Figure S5), the GDM accounted, on average across the 100 phylogenetic trees randomly resolved among congeneric species, for $6.6 \pm 0.3\%$ and $11.5 \pm 0.7\%$ of the total deviance in mosses and liverworts, respectively. In mosses, the best predictor was RH, which contributed to more than 99% of the explained deviance. A different pattern was observed in liverworts, where the variable accounting for within versus among trunk/canopy comparisons accounted for $65.6 \pm 3.3\%$ of the explained deviance, while the position of each individual tree in the x-y space and RH each contributed to less than 20% of the explained deviance.

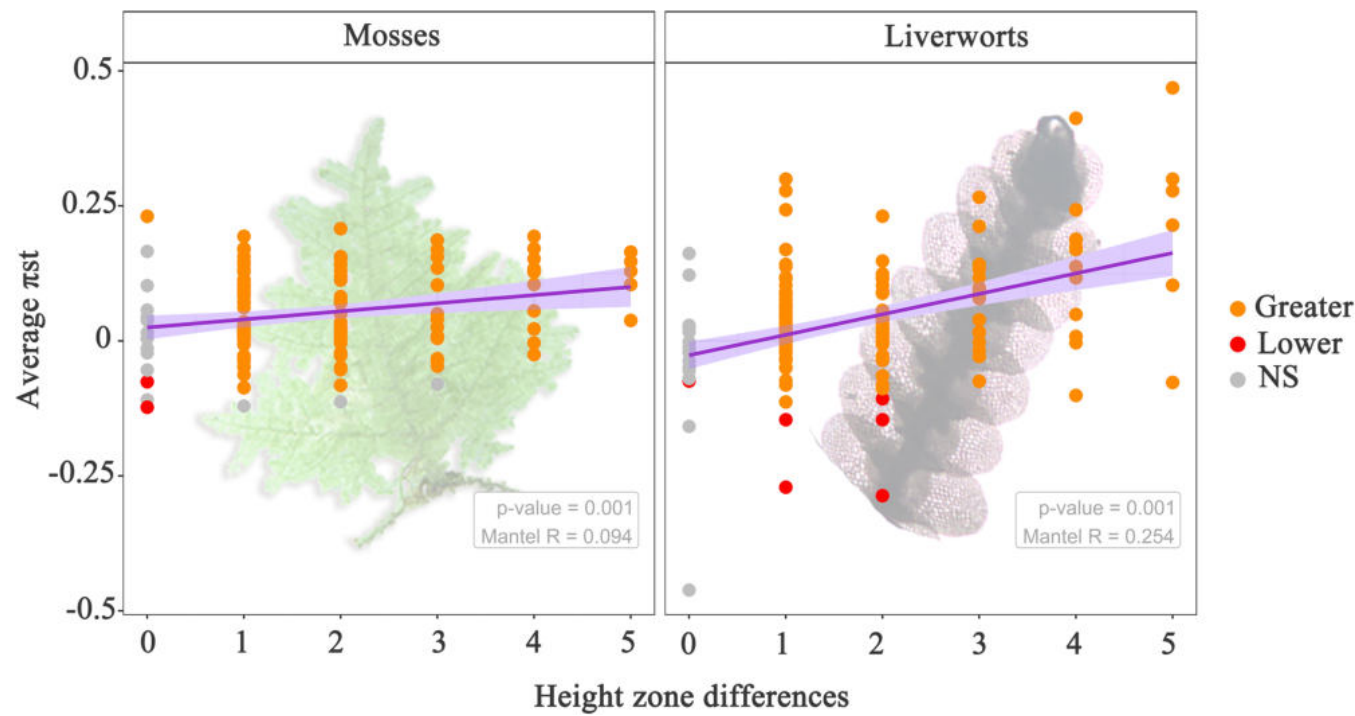

**FIGURE 5** Variation of average phylogenetic turnover within epiphytic moss and liverwort communities from the same height zone and on trees from the same DBH class in a lowland dipterocarp forest (Xishuangbanna, Yunnan, SW China) along a gradient of height zone difference. Average πst for a given environmental distance class that were significantly greater, significantly lower, and non-significantly (NS) different than expected by chance, are represented by orange, red and grey dots, respectively.

## 4 | DISCUSSION

The microclimatic data reported here and their three-dimension modelling help to fill a gap in our knowledge of the spatial variation of forest microclimates and to address the question of how microclimatic variation within and below tree canopies impacts community species richness and composition (De Frenne et al., 2019; Ellis & Eaton, 2021; Murakami et al., 2022). In line with previous assessments of microclimatic variation within canopies (see Murakami et al., 2022 and references therein), variations in temperature and RH were progressively buffered from the canopy to the tree base, with a substantially higher day/night difference in temperature and RH in the canopy. Monthly average temperatures of 28.9°C were recorded in the upper canopy and 26.1°C at the base, in line with previous reports of a mean difference of 4°C between forest understory and open ground due to the absorption of solar radiation by the canopy and increased evapotranspirative cooling (De Frenne et al., 2019).

The much wider range of vertical versus horizontal variation in microclimatic conditions explains why vertical species turnover is significantly higher than horizontal species turnover, despite the large differences in habitat conditions between young and old trees in terms of bark texture and chemistry (Fritz & Heilmann-Clausen, 2010; Fritz, Niklasson, & Churski, 2009; Wagner et al., 2015) as well as the effect of time, reflected by age differences among host trees, which impacts on the likelihood of colonization (Hidasi-Neto et al., 2019). Due to the prevalence of vertical microclimatic gradients, the contribution of nestedness to vertical beta diversity is negligible because specialist species segregate among height zones. In Amazonian rainforests for instance, more than half of the epiphyte species are height-zone specialists (Mota de Oliveira et al., 2009). Consequently, communities from the canopy share almost no species with communities from the tree base, preventing any nested pattern from emerging.

Our models accounted for 33% and 18% of the variation in vertical turnover in mosses and liverworts, respectively, and the predominant contributions of microclimatic factors (83–98%) to this pattern evidences their crucial role in determining the composition of epiphytic communities. The similar explanatory power of microclimatic conditions for moss and liverwort species turnover hides, however, opposite patterns of species richness in the two groups, with moss richness decreasing and liverwort richness increasing from the base to the canopy. Horizontal turnover in the two groups was similarly explained by the same factor, that is, tree size, but moss species richness peaked on large trees, while liverwort species richness peaked on small trees. Epiphytic moss and liverwort community composition thus responded in an opposite way to the same gradients, highlighting substantial differences in niche preferences between them. In vascular epiphytes, large, old trees tend to host a higher epiphytic richness than young ones due to the larger amount of time for colonization, larger space availability and greater diversity of microhabitats (Mayumi Francisco et al., 2021; Patiño et al., 2018; Zotz & Schultz, 2008; Zotz & Vollrath, 2003), and have therefore been a major focus for conservation (Adhikari et al., 2021; Wang et al., 2017). Our results suggest that epiphytic bryophyte diversity assessments in tropical forests must also include small, understorey trees (Sporn et al., 2010), which should also be considered for conservation.

Despite a comprehensive set of environmental variables, our models accounted for only 1/5–1/3 of the variation in species turnover, within the range of similar analyses for vascular epiphytes (0.10–0.57, Zotz & Schultz, 2008; Woods et al., 2015). Although additional variables characterizing microhabitat conditions, such as bark texture and chemistry, branch diameter, or percentage cover of canopy humus (Woods et al., 2015), would certainly increase model accuracy, we interpret the large proportion of unexplained variance in terms of stochasticity associated with dispersal limitations. Although epiphytes need to track patches of suitable trees (or leaves in the case of epiphylls) in a dynamic landscape for persistence (Snäll et al., 2005), mounting evidence suggests that dispersal capacity is counter-selected in epiphytic bryophytes. Epiphytic bryophytes typically exhibit spatially clustered distributions (Löbel et al., 2006; Snäll et al., 2003; Wagner et al., 2015) and their fine-scale patterns of genetic variation are strongly spatially structured (Ledent et al., 2020; Vanderpoorten et al., 2019), pointing to important effects of isolation-by-distance. These patterns are paralleled by morphological adaptations counter-favouring dispersal. For instance, the peristome, a ring of hygroscopic teeth that enhance spore dispersal in mosses, and the seta, which elevates the capsule above the substratum, are typically reduced in epiphytic species (Hedenäs, 2012). Peristome reduction is itself significantly associated with hygrochasy, i.e., the release of spores under wet conditions (Zanatta et al., 2018), further decreasing chances of long-distance dispersal but enhancing rates of establishment (Johansson et al., 2016).

The shift between negative or non-significant average phylogenetic turnover to consistently significant clustering that was observed along a gradient of height zone differences suggests that phylogenetic constraints further contribute to shaping the assembly of epiphytic bryophyte communities. The slight, but significant correlation between this trend for an increasing phylogenetic clustering with variation in microclimatic conditions adds to emerging evidence for the role of phylogenetic niche conservatism in community assembly through time (Saladin et al., 2019; Segovia et al., 2020), including at the much smaller spatial scales of epiphytic communities. Phylogenetic niche conservatism in epiphytic bryophyte communities, along with mounting evidence for niche conservatism in vascular epiphytes (Müller et al., 2017), shows that the specialization for vertical niches and their associated microclimatic conditions is phylogenetically inheritable, and hence, that species may be limited in their ability to shift among niches. The deep phylogenetic level (genus-level phylogeny), at which the analysis was conducted, further points to deeply nested phylogenetic constraints, which may have evolved during the burst of diversification of epiphytic lineages triggered by the development of large, humid, megathermal angiosperm forests (Feldberg et al., 2014).

The fact that there was no significant horizontal phylogenetic clustering of liverworts communities, and that the horizontal phylogenetic clustering observed in moss communities was not explained by differences in DBH among trees, conversely suggests that the succession of communities on a tree depending on its age is not phylogenetically constrained. Typically, early pioneers are short-lived organisms with a high reproductive effort, whereas late-colonizers have a longer lifespan and are characterized by limited reproductive investment (During, 1992). Although restricted to a set of 42 trees in a specific 1.44 ha plot, our results thus suggest that these life-history strategies arose multiple times during the evolutionary history of epiphytes.

The negative πst observed at tree base in mosses and liverworts is consistent with the expectation that competition in epiphytic communities should occur at levels characterized by lower variation of daily and seasonal microclimatic conditions rather than high-up in the canopy (Spicer & Woods, 2022). Although even strong competition levels can leave no trace in community phylogenetic structure (Bennett et al., 2013), the non-significant phylogenetic turnover that mostly characterized communities from the same height zone and trees of the same DBH class does not support the idea that competition plays an important role in shaping epiphytic bryophyte communities. In vascular epiphytes of lowland rain forests, which use only a small proportion of the available bark surface, the importance of competition has been similarly questioned (Zotz, 2016; Zotz & Vollrath, 2003). Competition could, however, be more important in montane forests, where epiphytes are typically much more abundant (Burns & Zotz, 2010). Instances of niche displacement were already reported among epiphytic bryophytes (Wiklund & Rydin, 2004), raising a series of questions on how species may shift niche to avoid competition.

## 5 | CONCLUSIONS

We provide here, through the spatially-explicit modelling of microclimatic conditions in a tropical forest, explicit support for the long-held notion that vertical variation in light, temperature and humidity conditions are the main driver of epiphytic species turnover along a tree. Epiphytic bryophyte communities were phylogenetically clustered, and the low, but significant correlation between phylogenetic turnover among communities and vertical microclimatic variation evidences fine-scale phylogenetic niche conservatism. Despite the comprehensive description of the host-tree environment, our analyses captured, however, only 1/5–1/3 of the floristic variation among communities, calling for further improvements and opening the door to new research perspectives. First, our analyses did not allow us to assess the potential role of positive interactions. In vascular epiphytes, positive co-occurrence patterns suggest potential facilitation (Burns & Zotz, 2010; Ceballos et al., 2016), as a dense clumping of epiphytes could enhance temperature and drought stress, which is expected to increase towards the outer canopy (Spicer & Woods, 2022). Second, our analyses failed to consider bryophyte-vascular epiphyte interactions, whose role in the structuring of bryophyte communities should be further investigated as vascular and bryophytic epiphytes significantly co-occur (Lu et al., 2020; Zotz & Vollrath, 2003).

The tight link between community composition and microclimatic conditions, as well as evidence for niche conservatism, raise questions about the ability of epiphytic bryophyte communities to move down along the trunk to track the shift of their niche in the context of climate change. How macroclimatic changes will impact the changes within canopies remains, however, uncertain. While the statistical modelling of microclimatic conditions as we implemented here may successfully capture the spatial variation of microclimatic conditions, the potential of such approaches to forecast novel conditions is somewhat questionable, calling for the development of mechanistic models based on first-principles physics (Maclean & Klinges, 2021).

## AUTHOR CONTRIBUTIONS

A.V., T.S., L.S., R.T.C. conceived the study. T.S., L.S., W.-Z.M., J.W., Y.-M.W., L.M. and Y.L. collected the data. T.S., F.C., T.K., Y.S. and L.M. performed the analyses. F.C., X.L.G., J.P. and O.H. provided suggestions throughout the process. A.V., T.S., R.T.C. and L.S. wrote the paper with the assistance from all co-authors.

## ACKNOWLEDGEMENTS

We thank Wen Yang for assistance with designing charts, Hai-Long Zhang and Jin-Long Dong for their assistance with field work, Simon Ferrier for assistance with the GDM model and the Xishuangbanna Tropical Rainforest Ecosystem Station (XTRES) of Chinese Ecosystem Research Network (CERN) and Sino BON-Forest Canopy Biodiversity Monitoring Network for providing environmental data. This study was funded by the National Natural Science Foundation of China (32171529), the Natural Science Foundation of Yunnan Province (202101AT070059), the CAS "Light of West China" Program, the CAS 135 program (No. 2017XTBG-F03, 2017XTBG-F01), the candidates of the Young and Middle-Aged Academic Leaders of Yunnan Province (2019HB040), and the Yunnan High Level Talents Special Support Plan (YNWR-QNBJ-2020-066). T.S. is funded by China Scholarship Council (No. 201904910636). J.P. is funded by the Ministerio de Ciencia e Innovación (MICINN) through the Ramón y Cajal program (RYC-2016-20506) and the grant (ASTERALIEN - PID2019-110538GA-I00) and by the Fundación BBVA (INVASION - PR19_ECO_0046). O.H. and A.V. are research directors of the Belgian Funds for Scientific Research (FRS-FNRS). Computational resources were provided by the Fédération Wallonie-Bruxelles (Tier-1; funded by Walloon Region, grant no. 1117545), and the Consortium des Équipements de Calcul Intensif (CÉCI; funded by the F.R.S.-FNRS, grant no. 2.5020.11).

## CONFLICT OF INTEREST

The authors declare no conflict of interest.

## PEER REVIEW

The peer review history for this article is available at https://publons.com/publon/10.1111/1365-2745.14011.

## DATA AVAILABILITY STATEMENT

The raw information on each individual tree and epiphytic species distribution data are available at https://doi.org/10.6084/m9.figshare.17057615.v8 (Shen, 2021a). The microclimatic data are available at https://doi.org/10.6084/m9.figshare.17057624 (Shen, 2021b).

## ORCID

*Ting Shen*  https://orcid.org/0000-0002-3061-624X
*Richard T. Corlett*  https://orcid.org/0000-0002-2508-9465
*Flavien Collart*  https://orcid.org/0000-0002-4342-5848
*Xin-Lei Guo*  https://orcid.org/0000-0002-2066-5929
*Jairo Patiño*  https://orcid.org/0000-0001-5532-166X
*Yang Su*  https://orcid.org/0000-0002-4717-9971
*Olivier J. Hardy*  https://orcid.org/0000-0003-2052-1527
*Wen-Zhang Ma*  https://orcid.org/0000-0003-3144-001X
*Liang Song*  https://orcid.org/0000-0002-1452-9939
*Alain Vanderpoorten*  https://orcid.org/0000-0002-5918-7709

## SUPPORTING INFORMATION

Additional supporting information can be found online in the Supporting Information section at the end of this article.